\documentclass[
reprint,
superscriptaddress,
amsmath,amssymb,
aps,
pra,
]%
{revtex4-1} 

\usepackage[utf8]{inputenc}
\usepackage[T1]{fontenc}
\usepackage[english]{babel}
\usepackage[autostyle=true]{csquotes}
\usepackage[normalem]{ulem}
\usepackage{xcolor}
\usepackage{graphicx}
\usepackage{dcolumn}
\usepackage{bm}
\usepackage{hyperref}
\hypersetup{
	    pdftitle={Moment of Inertia and the Scissors Mode of a Supersolid  Dipolar gas},
	    pdfauthor={Sandro Stringari},
		colorlinks=true,
		allcolors=blue,
}

\usepackage[caption=false]{subfig}
\graphicspath{{figures/}}
\usepackage[separate-uncertainty=false]{siunitx}
\DeclareSIUnit\gauss{G}

\usepackage{braket}

\usepackage[capitalise]{cleveref}

\usepackage{xstring}
\newcommand*{\aref}[1]{%
	\IfBeginWith{#1}{eq:}{Eq.~\eqref{#1}}{}
	\IfBeginWith{#1}{fig:}{Fig.~\ref{#1}}{}%
	\IfBeginWith{#1}{tab:}{Table~\ref{#1}}{}%
	\IfBeginWith{#1}{appendix:}{Appendix~\ref{#1}}{}%
	\IfBeginWith{#1}{sec:}{Section~\ref{#1}}{}%
	}
	
\begin{document}

\title{  Moment of Inertia  and Dynamical Rotational Response of a Supersolid Dipolar Gas. }
\date{\today}
\author{S. M. Roccuzzo}
\affiliation{INO-CNR BEC Center and Dipartimento di Fisica, Universit\`a degli Studi di Trento, 38123 Povo, Italy}
\author{A. Recati}\email[]{alessio.recati@ino.it}
\author{S. Stringari}
\affiliation{INO-CNR BEC Center and Dipartimento di Fisica, Universit\`a degli Studi di Trento, 38123 Povo, Italy}
\affiliation{Trento  Institute  for  Fundamental  Physics  and  Applications,  INFN,  38123,  Trento,  Italy}

\begin{abstract}

We show that the knowledge of the time dependent response of a trapped gas, subject to a sudden rotation of a confining harmonic potential, allows for the determination of the moment of inertia of dipolar supersolid configurations. While in the presence of one-dimensional arrays of droplets  the frequency of the resulting scissors oscillation provides accurate access to the value of the moment of inertia, two-dimensional like configurations are characterized by a multi-frequency structure in the rotating signal, reflecting the presence of significant rigid body components in the rotational motion.  
Using the formalism of  response function theory and simulations based on the so-called extended time dependent Gross-Pitaevskii equation, we point out the crucial role played by the low frequency components in the determination of the moment of inertia and of its deviations from the irrotational value. 
We also propose a protocol based on the stationary rotation of the trap, followed by its sudden stop, which might provide a promising alternative to the experimental evaluation of the moment of inertia. 

\end{abstract}

\maketitle

\section{Introduction}
After the first realization of the supersolid phase of a dipolar atomic gas \cite{F1,I1,S1}, the possibility of exploiting the superfluid behavior of these novel configurations  has already stimulated several theoretical and experimental studies. In addition to the occurrence of the new Goldstone modes emerging from the spontaneous breaking of translational invariance \cite{AndreevLifshitz,Boninsegni_SS2012,Rica1994}(see \cite{Watanabe12} for a  more general discussion), demonstrated in recent experiments with dipolar gases \cite{F2,I2,S2}, supersolids   are characterized by peculiar rotational features, like the deviations of the moment of inertia from the rigid body value \cite{Leggett1970,Leggett1998, roccuzzo2020} -- as a consequence of their superfluid nature -- and the possibility for such systems to host quantized vortices \cite{gallemi2020}. The moment of inertia is expected to play a crucial role in characterizing the frequency of the so called scissors oscillation, following the sudden rotation of an axially deformed  confining trap,  and, indeed, the theoretical predictions of Ref. \cite{roccuzzo2020} were soon confirmed by the experiment of Ref. \cite{ScissorPisa}. 
The investigation of the scissors mode has been actually employed in the past in different contexts to reveal the superfluid behavior of harmonically confined Bose \cite{guery_odelin1999,marago2000}, and Fermi \cite{Wright07} gases as well as of self bound configurations, like atomic nuclei \cite{scissors_nuclei_1,scissors_nuclei_2,scissors_nuclei_3}, and single droplets of dipolar gases \cite{Barbut18}. The basic idea of these works is the existence of a direct correspondence between the moment of inertia and the frequency of the scissors mode, the moment of inertia providing the collective mass parameter associated with the oscillation. Similar ideas were also used to investigate the  deviations of the moment of inertia from the rigid body value  in the low temperature phase of solid helium, employing a torsion oscillator approach \cite{Chan1,Chan2}. The experiments on solid helium have not however so far provided any evidence for supersolidity. 

In the presence of  supersolid configurations a simple relationship between the moment of inertia and the frequency of the scissors mode  is not however trivially ensured, because in general one expects that  the dynamic behavior of such systems is the result of a more complex combination of the rigid body motion of the droplets and of the irrotational flow of the surrounding superfluid gas, resulting in the emergence of multi  frequency components in the excitation spectrum of the rotating oscillation.  In the case of the highly elongated configurations considered in Refs. \cite{roccuzzo2020,ScissorPisa} it was shown that the dominant frequency, resulting from the sudden rotation of the trap, is directly related to the value of the moment of inertia. In a very recent experiment carried out in Innsbruck \cite{IBKscissor2021}, characterized by  more two-dimensional configurations, it was shown that such a simple relationship is not longer applicable.

The main purpose of the present work is to shed light into the dynamic behavior of the scissors mode of such  configurations using a sum rule approach and the formalism of linear response theory, going beyond the single mode approximation, as well as taking advantage of numerical simulations based on the solution of the extended Gross-Pitaevskii equation.

\section{Sum rules and linear response approach}
Trapped dipolar atomic gases of atoms with mass $M$ are described by a Hamiltonian $H$, which contains the single particle harmonic trapping potential $V_{ho}=M/2(\omega_x^2x^2+\omega_y^2y^2+\omega_z^2z^2)$, the two-body short range contact potential $V_{g}={4\pi\hbar a\over M}\delta({\bf{x}})$ characterised by the $s$-wave scattering length $a$ as well as the dipole-dipole interaction between
two identical magnetic dipoles $\mu$ aligned along the $z$ axis: $V_{dd}({\bf r}) =
\frac{\mu_0\mu^2}{4\pi}\frac{1-3\cos^2\theta}{|{\bf r}|^3}$  ($\theta$ being the angle between ${\bf
r}$ and the $z$ axis). In order to estimate the relative strength of the dipolar and the contact interaction, it is useful to define the adimensional parameter 
$
    \epsilon_{dd}=a_{dd}/a,
$
with $a_{dd}=\frac{\mu_0\mu^2}{12\pi\hbar}$ the characteristic dipolar length.   

The staring point of our discussion is the commutation relation 
\begin{equation}
[H,L_z]= i \hbar M(\omega^2_y-\omega^2_x)Q \; ,
\label{eq:commutation}
\end{equation}
which shows that there exists a direct coupling between the angular momentum $L_z=\sum_i (x_ip_i^y-y_ip_i^x)$ and the quadrupole $Q= \sum_i x_iy_i$   operators, where the sums run over all the $N$ atoms forming the gas. The coupling is  caused by the deformation of the harmonic trap and plays a crucial role in the study of the excitation spectrum. 

A first useful result obtainable from  \aref{eq:commutation} is the derivation of  a rigorous upper bound to the frequency of the lowest energy state excited by the angular momentum operator in terms of the energy weighted  and inverse energy weighted  moments of the zero-temperature angular momentum  dynamic structure factor $S_{L_zL_z}(\omega)= \sum_n |\langle 0|L_z|n\rangle|^2\delta(\hbar\omega-\hbar\omega_{n0})$, where $|0\rangle$ and $|n\rangle$ are the ground state and excited states of the many-body system, respectively, and $
\hbar\omega_{n0}$ is the energy difference between them. The upper bound can be written as \cite{roccuzzo2020}
\begin{equation}
\omega^2_{1,-1}\equiv \frac{m_1(L_z)}{m_{-1}(L_z)}= \frac{M(\omega^2_y-\omega^2_x)\langle x^2-y^2\rangle}{\Theta}
\label{eq:santo}
\end{equation}
where $m_p(L_z)=\int d\omega \omega^p S_{L_zL_z}(\omega)$ are the moments of the angular momentum dynamic structure factor and we used the sum rule result $m_1(L_z)= \langle 0|[L_z,[H,L_z]] |0\rangle /(2\hbar^2 )= N M (\omega^2_y-\omega^2_x)\langle x^2-y^2\rangle/2$  for the energy weighted moment. Here and in the following we use the notation $\langle f({\bf{r}})\rangle=\int d{\bf{r}}f({\bf{r}}) n({\bf{r}})$ with $n(r)$  the density distribution of the atomic cloud.  The inverse energy weighted moment is instead directly related to the moment of inertia, according to $\Theta =2\int d\omega S_{L_zL_z}(\omega)/\omega$, following from the static linear response  $N\Theta = \lim_{\Omega \to 0} \langle 0| L_z |0\rangle / \Omega$ of the system  to a  perturbation of the form $-\Omega L_z$.
The upper bound \aref{eq:santo} was shown in \cite{roccuzzo2020} to provide an excellent estimate of the scissors frequency in supersolid configurations characterized by a pronounced one-dimensional droplet alignment. For such configurations the experimental measurement of the scissors frequency then provides direct access to the value of the moment of inertia \cite{ScissorPisa}.

Another estimate of the scissors frequency can be derived using  the ratio between the cubic and the energy weighted moments of the dynamic structure factor relative to the angular momentum operator. Using the sum rule   $m_3= \langle 0| [ [L_z,H],[H,[H,L_z]]]|0\rangle/(2 \hbar^4) = N M (\omega^2_y-\omega^2_x)^2\langle x^2+y^2 \rangle /2$  one finds the result
\begin{align}
\omega^2_{3,1}\equiv\frac{m_3(L_z)}{m_{1}(L_z)}= & (\omega^2_y-\omega^2_x)\frac{ \langle x^2+y^2\rangle}{\langle x^2-y^2\rangle}= \nonumber \\
&\frac{M(\omega^2_y-\omega^2_x)\langle x^2-y^2\rangle}{\Theta_{irr}}
\label{eq:m3m1}
\end{align}
where, in the last equality, we have introduced the variational irrotational value of the moment of inertia defined by \begin{equation} 
\Theta_{irr}=  \beta^2\Theta_{rig} 
\label{Thetairr}
\end{equation}
with $ \beta= \langle x^2-y^2\rangle/\langle x^2+y^2\rangle$ the deformation of the atomic cloud and 
$
\Theta_{rig}= M \langle x^2+y^2\rangle
$ the rigid body value of the moment of inertia.  
It is worth noticing that the ratio \aref{eq:m3m1}, differently from \aref{eq:santo}, does not depend on the actual value of the moment of inertia, but only on the deformation factor $\beta$ of the atomic cloud.

Being based on different moments of the dynamic structure factor, the two estimates satisfy the rigorous inequality $\omega_{1,-1}\le \omega_{3,1}$ and coincide only if the moment of inertia is equal to  the irrotational value. This corresponds to the condition that a single excited state exhausts the moments $m_p(L_z)$.
 The condition $\omega_{1,-1}\le \omega_{3,1}$ is consistent with the fact that $\Theta_{irr}$  provides a variational estimate of the moment of inertia, obtained by imposing the irrotational velocity flow ${\bf v}= \alpha \nabla xy$ in the calculation of the response to the angular momentum constraint $-\Omega L_z$. The quantity  $\Theta_{irr}$ then provides   a lower bound to the actual value of the moment of inertia, which consequently satisfies the general condition
$\Theta_{irr} \le \Theta \le \Theta_{rig}$. In usual superfluid gases interacting with zero range interactions the identity $\Theta= \Theta_{irr}$ follows from the irrotational hydrodynamic theory of superfluids applied to harmonically trapped gases \cite{BecBook2016},  which actually predicts the well known result $\omega_{scissors}= \sqrt{\omega_x^2+\omega_y^2}$ for the scissors frequency \cite{guery_odelin1999}. The above result for the scissors frequency is  directly derivable from Eq.(\ref{eq:m3m1}) employing the  Thomas-Fermi approximation for the calculation of $\langle x^2-y^2\rangle$.  It is also worth noticing that the estimates \aref{eq:santo}  and \aref{eq:m3m1}  provide rigorous upper and lower bounds to the frequencies  of the lowest and highest energy states  excited by the operator $L_z$, respectively. 

\begin{figure}
    \centering
    \includegraphics[width=0.5\textwidth]{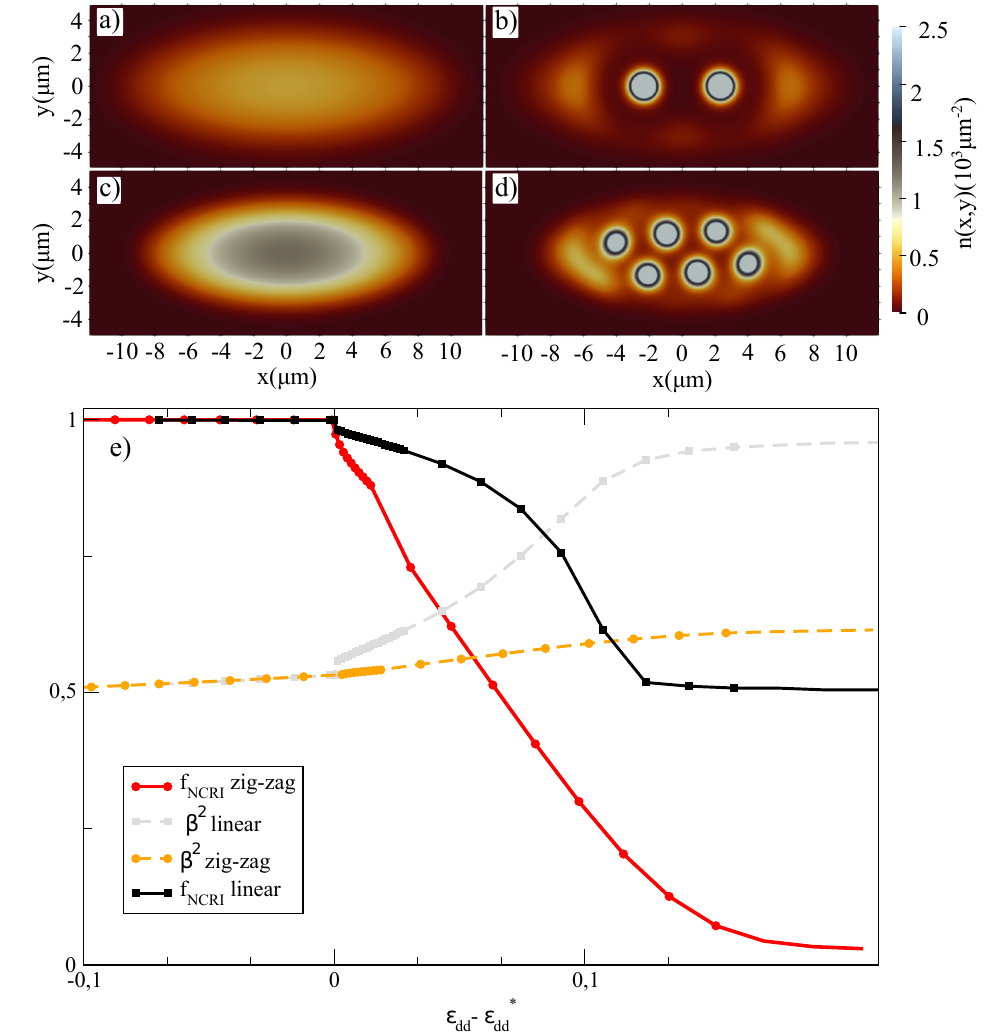}
\caption{ (a-d) Examples of integrated ground-state density profiles $n(x,y)=\int dz \Psi(x,y,z)|^2$, in the superfluid and supersolid phase, for gas of $^{164}$Dy atoms. Panels (a) and (b) refer to a gas of $40000$ atoms in a harmonic trap of frequencies $\omega_{x,y,z}=2\pi(20,40,80)Hz$, and $\epsilon_{dd}=1.32$ and $1.434$ respectively. Panels (c) and (d)  refer to a gas of $60000$ atoms in a harmonic trap of frequencies $\omega_{x,y,z}=2\pi(40,80,160)Hz$, and $\epsilon_{dd}=1.32$ and $1.467$ respectively. e) Fraction of non-classical rotational inertia, $f_{NCRI}$ Eq. \ref{fNCRI} (symbols and solid lines) and the ratio $\beta^2=\Theta_{irr}/\Theta_{rig}$ as function of $\epsilon_{dd}$ (symobols and dotted lines) for the "zig-zag" (red circles) configuration \cite{IBKscissor2021} and for the small linear configuration (black squares) considered in \cite{roccuzzo2020,ScissorPisa}. $\epsilon_{dd}^*$ is the value of $\epsilon_{dd}$ at which the superfluid-supersolid phase transition occurs.}
    \label{fig:fncri}
\end{figure}

The deviations of the actual value of the moment of inertia from the irrotational   and    rigid body   values can be  employed to calculate the so-called non classical rotational inertia (NCRI) fraction $f_{NCRI}$ \cite{Leggett1970,Leggett1998}, according to the relation \cite{Hydro-Rica2007,ScissorPisa}
\begin{equation}
f_{NCRI}= \frac{1-\Theta/\Theta_{rig}}{1-\Theta_{irr}/\Theta_{rig}} \; ,
\label{fNCRI}
\end{equation}
which satisfies the physical conditions of vanishing  if $\Theta=\Theta_{rig}$ and of being equal to unity if $\Theta=\Theta_{irr}$. The quantity $f_{NCRI}$ provides  a useful estimate  of the global superfluid character  of the system.
In \aref{fig:fncri} we show  $f_{NCRI}$ as a function of 
the relevant dimensionless $\epsilon_{dd}=a_{dd}/a$ interaction parameter across the superfluid/supersolid transition. 
Results are shown for the geometry considered in \cite{roccuzzo2020}, corresponding to
$N=4\times 10^4$ atoms of $^{164}$Dy in a harmonic trap of frequencies $\omega_{x,y,z}=2\pi(20,40,80)Hz$ (black squares), and for a geometry similar to the one considered in \cite{IBKscissor2021}, i.e., $N=6\times 10^4$ and $\omega_{x,y,z}=2\pi(40,80,160)Hz$ (red circles), in which the density peaks in the supersolid phase arrange in a "zig-zag" structure. Typical density configurations for the former (latter) cases in the superfluid and supersolid phase are shown in panels a) and b) (c) and d)) of \aref{fig:fncri}. 

It is worth noticing that the inter droplet distance is  proportional to  $\propto 1/\sqrt{\omega_z}$ \cite{SantosRoton} and is hence larger in panel b)  with respect to panel d).  The atomic configurations and the moments of inertia entering the function $f_{NCRI}$ were calculated by means of the so-called extended Gross-Pitaevskii equation as explained in the Appendix. 
The figure shows that relevant information on the amount of superfluidity carried by the system  is ensured by  both configurations,  the "zig-zag" configuration turning out to be  more sensitive to the presence of non superfluid components, as a consequence of the larger number of droplets and their more two-dimensional arrangement. As explicitly pointed out in \cite{roccuzzo2020,gallemi2020} the function $f_{NCRI}$ remains finite also for large values of $\epsilon_{dd}$, when the system is in the crystal phase of incoherent droplets, the superfluidity of the system being entirely due, in this case, by the internal motion of each droplet which cannot sustain a rigid rotational motion.  The effect is particularly important in the case of a linear droplet array. 
Let us also point out that a safe determination of the ratio Eq. \ref{fNCRI} requires that  $\Theta_{irr}$ be not too close to $\Theta_{rig}$. This is not an obvious requirement because the value of the cloud deformation  $\beta$ is not uniquely fixed by the aspect ratio of the trap, but is known to increase significantly with $\epsilon_{dd}$. For completeness in \aref{fig:fncri} we also report $\beta^2=\Theta_{irr}/\Theta_{rig}$.

In the theoretical calculation of \cite{roccuzzo2020}, carried out on a very elongated geometry, it was found that Eq. (\ref{eq:santo}) provides a very accurate estimate of  the frequency of the scissors mode resulting from the sudden rotation of the confining trap, thereby suggesting that the experimental  measurement of the scissors frequency can be used to infer the value of the moment of inertia. This was actually the procedure followed in \cite{ScissorPisa}. 
As explicitly pointed out in \cite{IBKscissor2021} the situation is different in more two-dimensional configurations, characterized by a smaller deformation of the density profile.  In \cite{IBKscissor2021} it was actually found that the frequency of the dominant oscillation, following the sudden release of the trap,  is close to the irrotational value (\ref{eq:m3m1}) even for configurations where the moment of inertia  deviates from the irrotational value in a significant way, therefore raising the question whether the experimental measurement of the  scissors mode can in general provide   a measure of the moment of inertia and  of the superfluid nature of the system.

\begin{figure*}[t]
    \centering
    \includegraphics[width = \columnwidth]{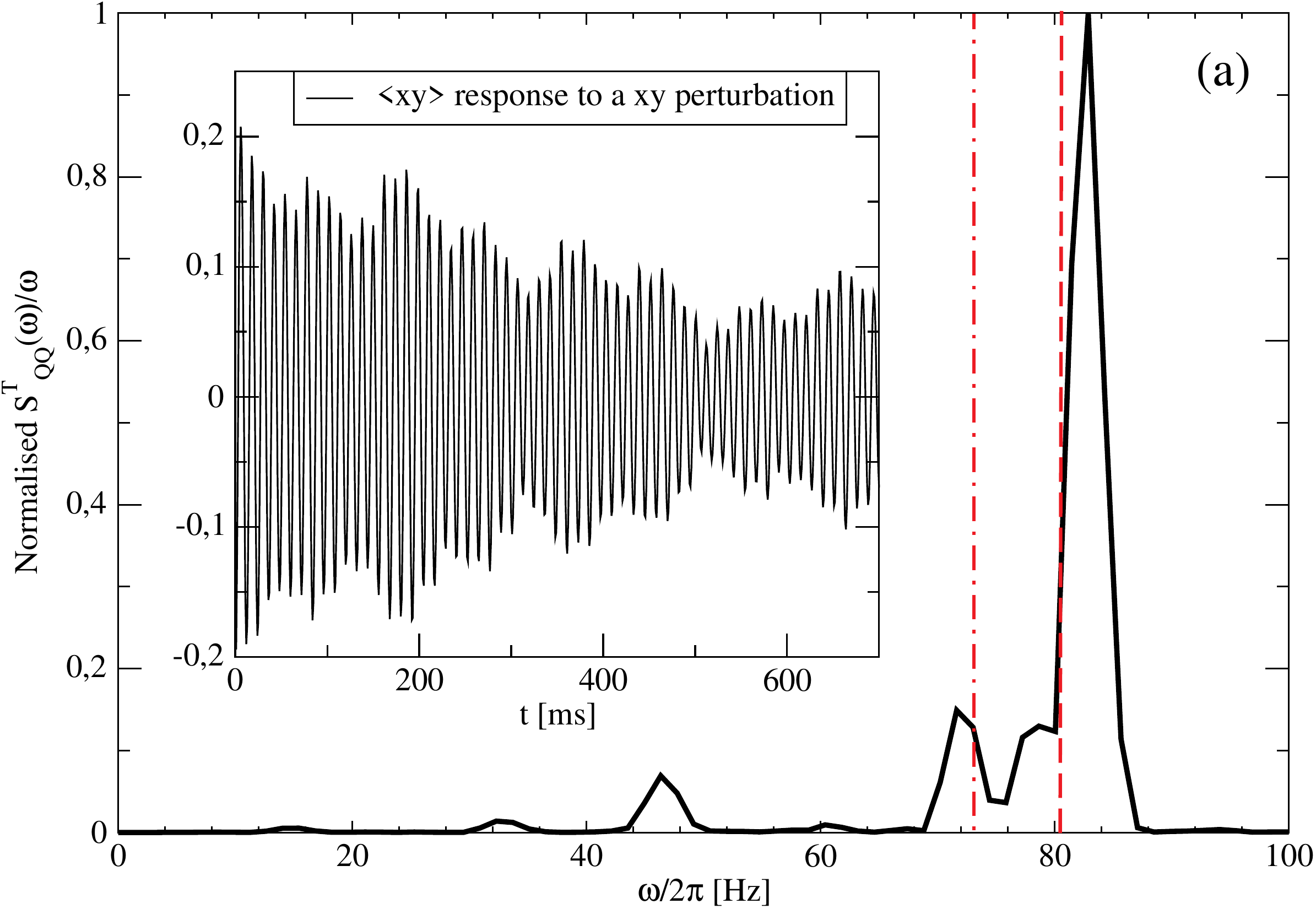}
   \includegraphics[width = \columnwidth]{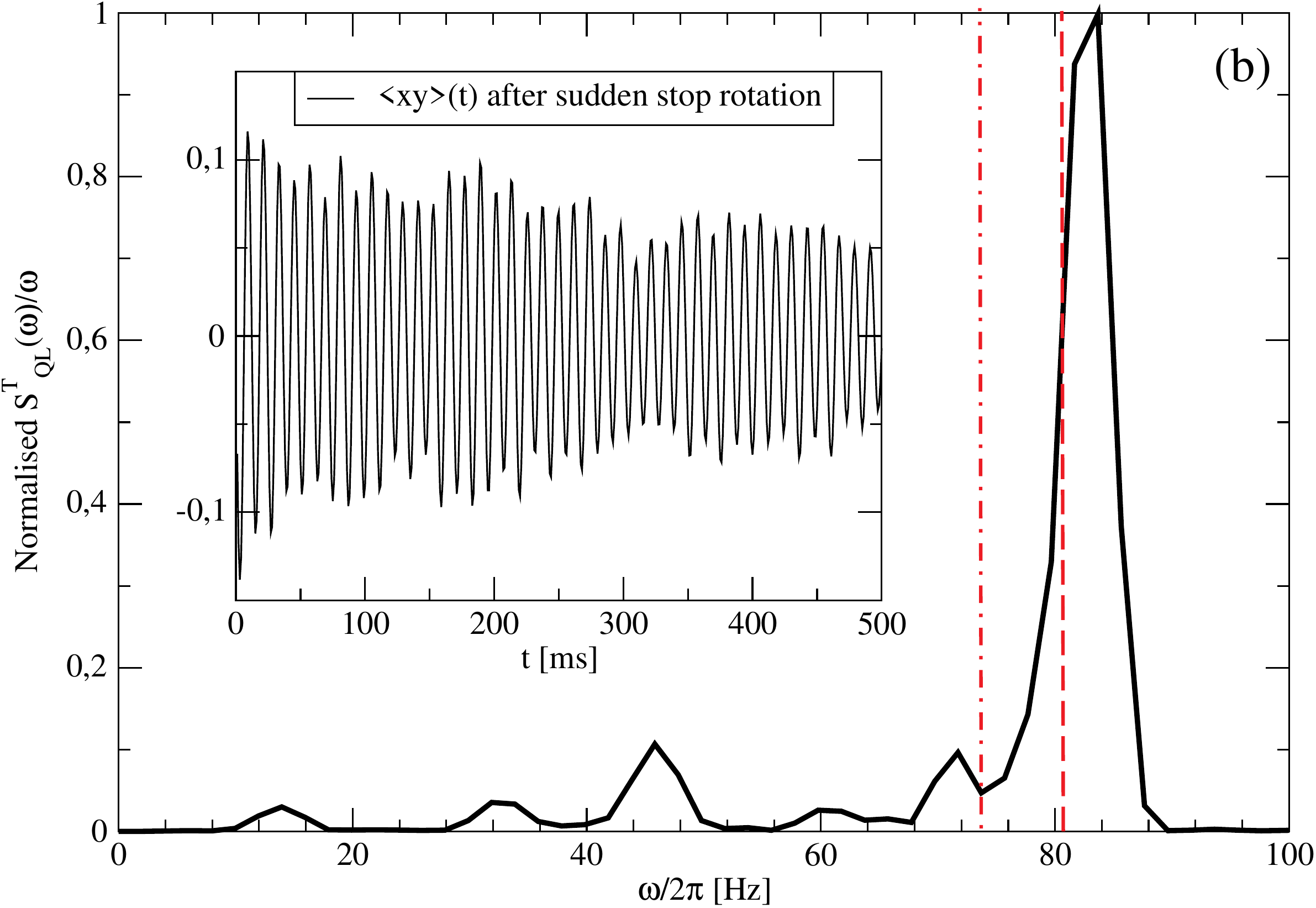}
   \includegraphics[width = \columnwidth]{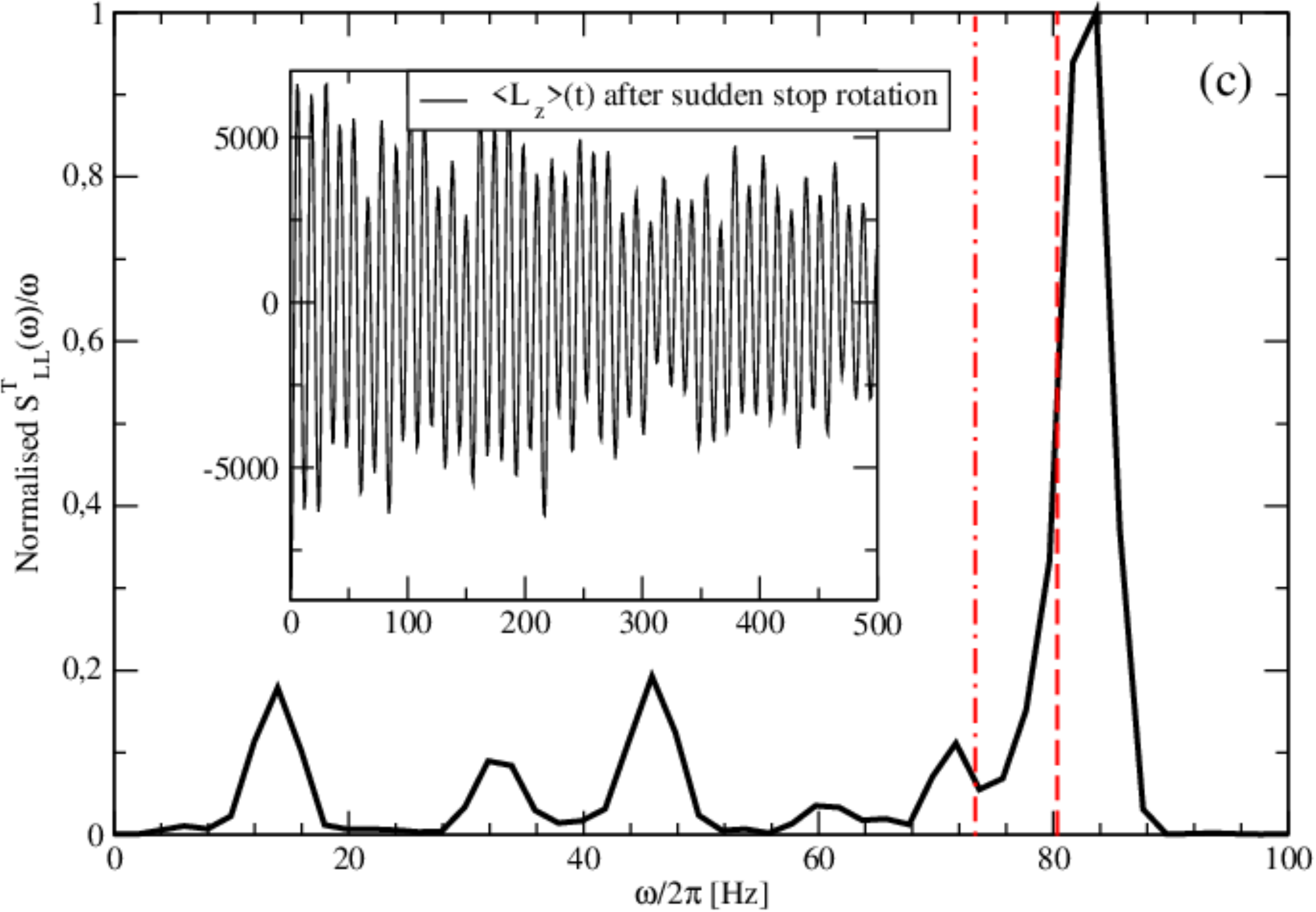}
\caption{
Time dependence (insets) and Fourier transform $S^T(\omega)/\omega$ (main panel) corresponding to different observables of the "zig-zag" configuration for $\epsilon_{dd}=1.467$:  (a) Response $\langle xy \rangle$  to a sudden trap rotation. (b) Response $\langle xy \rangle$  to a sudden stop of the trap rotation. (c) Response $\langle L_z \rangle$   to a sudden stop of the trap rotation. \\
    The vertical red dashed line is the value $\omega_{31}$ -- Eq. \ref{eq:m3m1} -- given by the irrotational moment of inertia, while the vertical red dash-dotted line is the value $\omega_{1,-1}$ -- Eq. \ref{eq:santo}-- given by the true moment of inertia of the system.}
    \label{fig:response}
\end{figure*}

In the following we will discuss how the moment of inertia can be in principle obtained from the knowledge  of the time dependence of the quadrupole moment $\langle xy \rangle (t)$, following the sudden rotation of the harmonic trap. This approach, first  applied to a Bose  gas below and above the BEC critical temperature \cite{guery_odelin1999,Zambelli01}, is applicable  also when the single mode approximation is not valid and the time dependence of  $\langle xy \rangle $ consequently exhibits a multi-frequency structure, with the appearence of low frequency components.  

The rotation of the trap in the $x-y$ plane by a small angle $\theta$, causes the appearence of a perturbation term  $\delta V_{ho}=\lambda xy$ with $\lambda = M \theta (\omega_y^2-\omega_x^2) $. Therefore we are left with the need to determine the response of $Q=xy$ due to a perturbation generated by the same operator. 
Using first order perturbation theory  one finds that, after the sudden rotation of the trap,  the time dependence of the signal can be written as  
\begin{equation}
\langle xy \rangle (t) = 2 \lambda \int d\omega \frac {S_{QQ}(\omega)}{\omega} \cos(\omega t) \; ,
\label{eq:xyxy}
\end{equation}
with the quadrupole dynamic structure factor $S_{QQ}(\omega)$ defined by $
S_{QQ}(\omega) = \sum_n |\langle 0|Q|n \rangle |^2 \delta(\hbar\omega-\hbar\omega_{n0})$. 
On the other hand, thanks to the commutation relation \aref{eq:commutation}, the quadrupole and the angular momentum dynamic structure factors are related to each other by the identity $S_{LL}(\omega) = \hbar M^2(\omega^2_y-\omega^2_x)^2S_{QQ}(\omega)/\omega^2$, so that the dynamic structure factor  $S_{QQ}(\omega)$, derivable  from the Fourier transform of the measured signal $\langle xy \rangle (t)$, allows for the determination of the moment of inertia according to \cite{Zambelli01}
\begin{equation}
\Theta = M(\omega^2_y-\omega^2_x)\langle x^2-y^2 \rangle\frac{ \int d\omega  S_{QQ}(\omega)/\omega^3}{\int d\omega S_{QQ}(\omega)/\omega} \; ,
\label{eq:TQQ}
\end{equation}
where we have further  used the identity 
\begin{eqnarray}
\int d\omega \frac{S_{QQ}(\omega)}{\omega} &=&{1\over \hbar M^2(\omega^2_y-\omega^2_x)^2}\int d\omega S_{L_zL_z}(\omega)\omega\\ \nonumber
&=& \frac{\hbar\langle x^2-y^2 \rangle}{2M(\omega^2_y-\omega^2_x)}
\end{eqnarray}

The simplest case of multi-frequency structure  is provided by the ideal classical gas, where the moment of inertia is  given by the rigid body value. In this case the response, in the collisionless regime, is characaterized by two frequencies   $|\omega_x\pm\omega_y|$ \cite{guery_odelin1999} and  the strengths of the two modes, fixing the time dependence of the physical signal  $\langle xy\rangle(t)$, are exactly equal, giving rise to a visible beating effect (see Fig. 3a in \cite{marago2000}). As we will discuss below, the situation is different in the case of the supersolid configurations considered in this work.   

\section{Numerical results}

Let us consider the supersolid "zig-zag" configuration shown in \aref{fig:fncri} (d) corresponfing to $\epsilon_{dd}=1.466$, which is close to the value considered in the recent experiment of \cite{IBKscissor2021}. In the inset of \aref{fig:response}(a-c) we show the time dependence of $\langle xy \rangle (t)$ for a time interval $t\in [0,T]$, following the sudden rotation of the trap. The moment of inertia of such a "zig-zag" configuration, determined by performing a stationary calculation of the eGP equation in the presence of the stationary constraint $-\Omega L_z$, is equal to  $0.66\,\Theta_{rig}$ while the irrotational value Eq. (\ref{Thetairr}) is given by  $0.55\,\Theta_{rig}$. As already discussed above, the difference between $\Theta$ and $\Theta_{irr}$ is an indicator of the presence of non superfluid effects in the rotational behavior.  The time dependent signal reveals, at short times, a fast oscillatory behavior at about 82 Hz, very close to the irrotational value Eq. (\ref{eq:m3m1}) -- the vertical red dashed line in \aref{fig:response}--, followed, at larger times, by beating effects signalling the occurrence of lower frequency components in the excitation spectrum. 
Assuming that the most relevant frequencies entering the dynamic structure factor $S_{QQ}(\omega)$ are larger than $2\pi/T$, we can use the estimate  $S_{QQ}(\omega)\simeq S^T_{QQ}(\omega)=\omega|FT[\langle xy\rangle]|(\omega)$, where $FT$ stands for the Fourier transform of the numerical signal. In the main panel of \aref{fig:response}(a) we show the extracted value  $S^T_{QQ}(\omega)/\omega$. 
Our analysis  shows that, although the signal is dominated by the high frequency mode with frequency close to the irrotational value $\omega_{31}$, the emergence of the  lower frequency components is  however crucial for the correct calculation of the relevant integral $\int d\omega S_{QQ}/\omega^3$, giving access to the moment of inertia. In particular, limiting the Fourier analysis of  the signal to  times of much shorter duration $T$, and therefore missing the contribution given by the lower frequency components,  one would extract a significantly smaller value of the moment of inertia, closer to the irrotational value. 
 The value of $\Theta$ extracted according to Eq.(\ref{eq:TQQ}) turns out to be $0.74\,\Theta_{rig}$,  pretty higher than the irrotational value, but still not yet enough accurate as compared to the exact value. The remaining discrepancy is not surprising since longer expectation  times in the evaluation of the signal $\langle xy \rangle $ would be required in order to get an accurate determination of the low frequency contribution to the relevant integral $\int d\omega S_{QQ}/\omega^3$.  In a real experiment this might require rather prohibitive conditions due to the short lifetime of the sample.

In order to amplify the role  of the low frequency modes in the observable signal,  an alternative procedure  consists of generating a steady rotation of the trap at a given angular velocity $\Omega$, followed by its sudden stop. In this case the measurable signal $\langle xy \rangle (t)$ obeys the law
\begin{equation}
\langle xy \rangle (t) = 2 \Omega \int d\omega \frac {S_{LQ}(\omega)}{\omega} \sin(\omega t) \; ,
\label{eq:xyLz}
\end{equation}
where we have defined the crossed angular momentum - quadrupole dynamic structure factor $S_{LQ}(\omega)= i\sum_n \langle 0|L_z|n \rangle \langle n|xy|0 \rangle  \delta(\omega-\omega_{n0})$ 
related to the angular momentum dynamic structure factor by the law $S_{LL}(\omega)=  (\omega^2_y-\omega^2_x)S_{LQ}(\omega)/\omega$, as a consequence of the commutation relation \aref{eq:commutation}. In this case the measurable signal $S_{LQ}(\omega)/\omega$ entering Eq.(\ref{eq:xyLz}) is more sensitive to the excitation of the low frequency modes and one  finally obtains  the result
\begin{equation}
\Theta = M(\omega^2_y-\omega^2_x)\langle x^2-y^2 \rangle \frac{ \int d\omega S_{LQ}(\omega)/\omega^2}{\int d\omega S_{LQ}(\omega)} 
\label{eq:TLQ}
\end{equation} 
for the moment of inertia. 
In \aref{fig:response}(b) we report the calculation of $\langle xy \rangle (t)$ and of its Fourier transform 
based on the numerical solution of the eGP equation. It is clear how the lower frequency signals are enhanced in this case. Following  this procedure we find the value  $0.69\,\Theta_{rig}$ for the moment of inertia,  closer to the exact value $0.66\,\Theta_{rig}$ obtained from the static eGPE calculation.

Even if not easily measurable, for completeness let us discuss the time dependence of the angular momentum $\langle L_z(t)\rangle$ carried by the system, after stopping the trap rotation. In this case the Fourier transform of the signal gives direct access to the angular momentum dynamic structure factor: 
\begin{equation}
\langle L_z\rangle(t) = 2 \Omega \int d\omega \frac {S_{LL}(\omega)}{\omega} \cos(\omega t) 
\label{eq:LzLz}
\end{equation}
and the moment of inertia can be written as 
\begin{equation}
\Theta = M(\omega^2_y-\omega^2_x)\langle x^2-y^2 \rangle \frac{ \int d\omega S_{LL}(\omega)/\omega}{\int d\omega \omega S_{LL}(\omega)}.
\label{eq:TLL}
\end{equation} 
The time dependence of $\langle L_z\rangle$ and its Fourier transform are reported in \aref{fig:response}(c). 
In this case the role of the low energy states, providing an important contribution to the moment of inertia sum rule (\ref{eq:TLL}),  emerges in an even  clearer  way and the  value of the moment of inertia, derivable from Eq. (\ref{eq:TLL}) turns out to be  $\Theta =0.67\,\Theta_{rig}$, in excellent agreement with the exact value. As already mentioned above, the experimental measurement of $L_z(t)$ is unfortunately not   available at present, but the identification of the low frequency states, responsible for their relevant contribution to the moment of inertia, could be obtained by applying a resonant modulation of the rotating trap at the proper frequency. This might also provide an alternative procedure  to measure directly the dynamic structure factor $S_{QQ}(\omega)$.

\begin{figure}
    \includegraphics[width = \columnwidth]{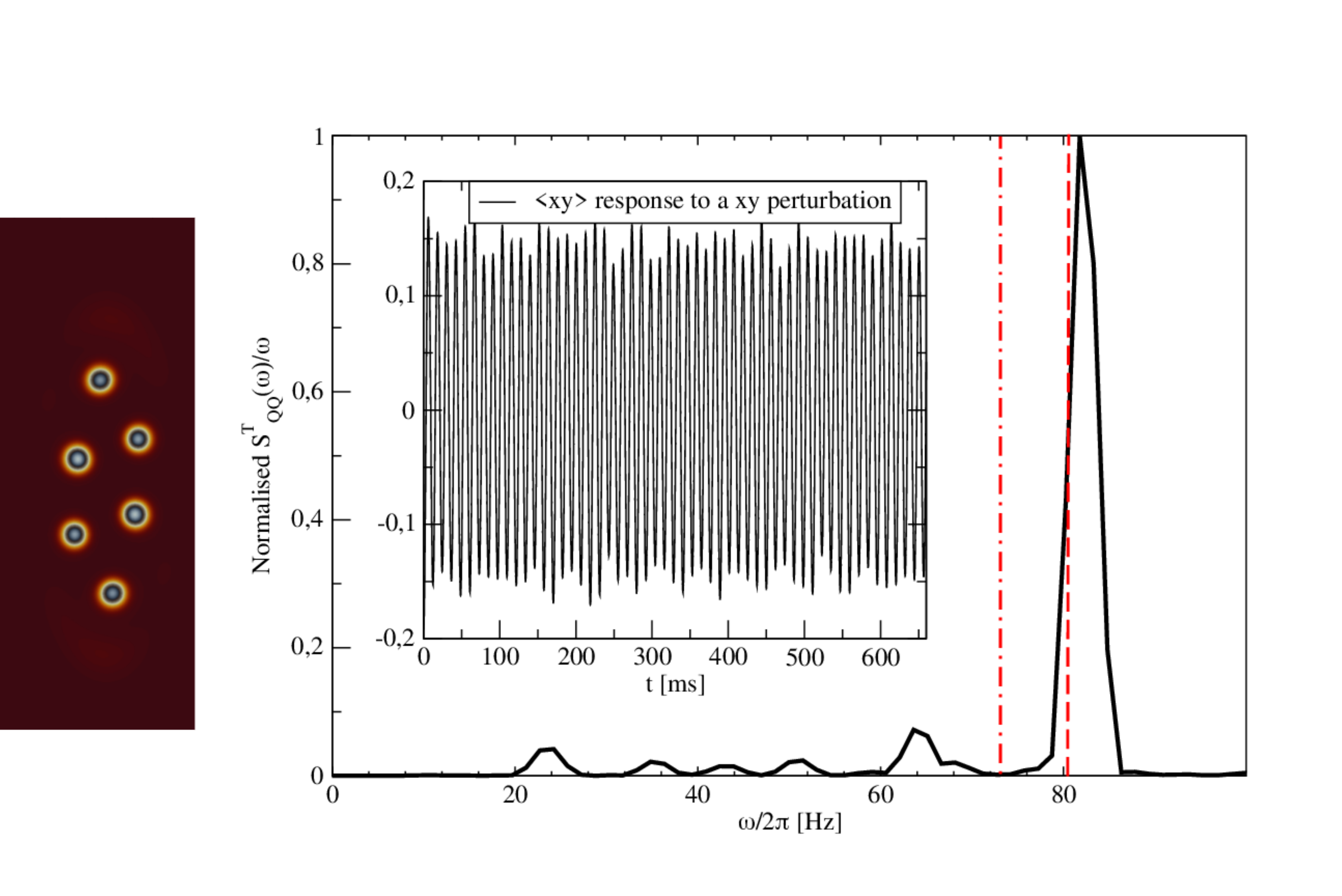}
    \caption{Left panel: Incoherent droplet configuration for the same parameters as in Fig. \ref{fig:fncri} panel (c) and (d), but $\epsilon_{dd}=1.553$. Right panel: response $\langle xy \rangle$ of the incoherent droplets to a sudden rotation of the trap.}
    \label{fig:ID}
\end{figure}
In \aref{fig:ID}  we finally report the calculated signal $\langle xy\rangle(t) $ and its Fourier transform in the case of a crystal of incoherent droplets, realized with a significantly  larger value of the interaction parameter ($\epsilon_{dd}=1.553$). In this case we find that, despite the almost rigid like behavior of the moment of inertia, the dynamic structure factor $S^T(\omega)/\omega$ is characterized by a dominant peak close to the irrotational value and a series of low frequency states providing the remaining contribution to the moment of inertia. We find that the moment of inertia, determined via \aref{eq:xyxy} using the correponding Fourier transform reported in \aref{fig:ID}, is in perfect agreement with the value $\Theta=0.91\Theta_{rig}$, determined by the static linear response. 

\section{Conclusions}

In conclusion, we have shown that the moment of inertia of a supersolid  dipolar gas confined by a harmonic trap can be determined through  the study of the time dependent response of the system to either the  sudden rotation of the trap or   the sudden stop of its stationary rotation. In both cases the Fourier transform of the observable signal $\langle xy \rangle(t)$ is shown to contain the relevant information needed to extract the value of the moment of inertia, provided the time duration of the oscillating signal is sufficiently long.  Our analysis  explicitly reveals    the superfluid character   of the  supersolid configurations whose moment of inertia differs from the rigid value,  but at the same time differs from the irrotational value, probing the occurrence of  solid like components in the rotational motion.  We have shown that, except for    1D aligned configurations, where the frequency of the  scissors mode is well approximated by the square root  ratio between the energy weighted and inverse energy weighted moments (see Eq. (\ref{eq:santo})), giving direct access to the moment of inertia, for more 2D like configurations the procedure to extract the moment of inertia requires a more careful analysis of the resulting signal, due the occurrence of the multi-frequency nature of the excitation spectrum, and consequently longer observation times. We have also shown that the procedure based on the sudden stop of the rotation of the trap would allow for an easier identification of the relevant low frequency components, which are
responsible for a significant contribution to the moment
of inertia. 

\section*{Acknowledgements}
We thank Giulio Biagioni, Francesca Ferlaino, Giovanni Modugno, Matthew Norcia and Elena Poli for insightful discussions.
This work was supported by Q@TN (the joint lab between University of Trento, FBK- Fondazione Bruno Kessler, INFN- National Institute for Nuclear Physics and CNR- National Research Council), the Provincia Autonoma di Trento and the Italian MUR under the PRIN2017 project CEnTraL (ProtocolNumber 20172H2SC4).

\appendix

\section{The extended Gross-Pitaevskii equation}
At zero temperature, a dipolar Bose gas can can be descibed in terms of a single complex order parameter $\Psi({\bf r},t)$, whose squared modulus gives the local density of the system, and whose temporal evolution is described by the following extended Gross-Pitaevskii equation:
\begin{eqnarray}
&&i\hbar\frac{\partial}{\partial t}  \Psi({\bf r},t)= 
 \left[ -\frac{\hbar^2}{2m}\nabla^2+g|\Psi({\bf r},t)|^2+\gamma(\varepsilon_{dd})|\Psi({\bf r},t)|^3\right. \nonumber \\
 &+& \left.\!
 \int V_{dd}({\bf r}-{\bf r'})|\Psi({\bf r'},t)|^2 d{\bf r'}+V_{\rm ho}({\bf r})\right] \Psi({\bf r},t)
 \label{appendix:egpe}
\end{eqnarray}
Here, $g=4\pi\hbar^2a/m$ is the coupling constant fixed by the $s$-wave scattering length $a$, $V_{\rm ho}({\bf r})$ is the external harmonic trapping potential and $V_{dd}({\bf r}_{i}-{\bf r}_{j})=\frac{\mu_0\mu^2}{4\pi}\frac{1-3\cos^2\theta}{|{\bf r}_{i}-{\bf r}_{j}|^3}$ is the dipole-dipole interaction potential, with $\mu_0$ the magnetic permeability of the vacuum, $\mu$ the magnetic dipole moment and $\theta$ the angle between the vector distance between dipoles and the polarization direction, which we choose as the $z$-axis. This model takes into account the effect of quantum fluctuations via the so-called Lee-Huang-Yang term, inserted in the local density approximation via the term proportional to $|\Psi({\bf r},t)|^3$. The prefactor $\gamma(\varepsilon_{dd})$ was first calculated in Ref. \cite{Pelster}, and for a homogeneous, three-dimensional dipolar gas is given by 
\begin{equation}
\gamma(\varepsilon_{dd})=\frac{16}{3\sqrt{\pi}} ga^{\frac{3}{2}}\,\mbox{Re}\bigg[\!\int_0^{\pi}\!\!\!\!d\theta\sin\theta [1+\varepsilon_{dd}(3\cos^2\theta-1)]^{\frac{5}{2}}\bigg]\,.
\end{equation}
where $\varepsilon_{dd}=\mu_0\mu^2m/12\pi\hbar^2a$ is an a-dimensional parameter measuring the relative strength of the contact and the dipole-dipole interaction. This coefficient can be adjusted in experiments using Feshbach resonances by tuning the value of the $s$-wave scattering length $a$. Experimental measurements and microscopic Monte Carlo calculations \cite{Saito2016}
have confirmed the accuracy of the extended Gross-Pitaevskii approach in taking into account  beyond mean field effects in both dipolar gases and quantum mixtures 
\cite{PetrovDrop,TarruellDrop,TarruellDrop2,FattoriDrop,FattoriColl}.

\bibliography{biblio.bib}

\end{document}